# Should citations be field-normalized in evaluative bibliometrics?

# An empirical analysis based on propensity score matching

Lutz Bornmann*, Robin Haunschild** & Rüdiger Mutz***

*Division for Science and Innovation Studies

Administrative Headquarters of the Max Planck Society

Hofgartenstr. 8,

80539 Munich, Germany.

E-mail: bornmann@gv.mpg.de

** Max Planck Institute for Solid State Research

Heisenbergstraße 1,

70569 Stuttgart, Germany.

Email: r.haunschild@fkf.mpg.de

*** Professorship for Social Psychology and Research on Higher Education

ETH Zurich

Andreasstrasse 15 (AND 2 80),

8050 Zurich, Switzerland.

Email: ruediger.mutz@gess.ethz.ch

**Abstract**

Field-normalization of citations is bibliometric standard. Despite the observed differences in citation counts between fields, the question remains how strong fields influence citation rates beyond the effect of attributes or factors possibly influencing citations (FICs). We considered several FICs such as number of pages and number of co-authors in this study. For example, fields differ in the mean number of co-authors (pages), and – on the paper level – the number of co-authors (pages) is related to citation counts. We wondered whether there is a separate field-effect besides other effects (e.g., from numbers of pages and co-authors). To find an answer on the question in this study, we applied inverse-probability of treatment weighting (IPW) which is a variant of the "propensity score matching" approach. Using Web of Science data (a sample of 308,231 articles), we investigated whether mean differences among subject categories in citation rates still remain, even if the subject categories are made comparable in the field-related attributes (e.g., comparable of co-authors, comparable number of pages) by IPW. In a diagnostic step of our statistical analyses, we considered propensity scores as covariates in regression analyses to examine whether the differences between the fields in FICs vanish. The results revealed that the differences did not completely vanish but were strongly reduced. We received similar results when we calculated mean value differences of the fields after IPW representing the causal or unconfounded field effects on citations. However, field differences in citation rates remain. The results point out that field-normalization seems to be a prerequisite for citation analysis and cannot be replaced by the consideration of any set of FICs in citation analyses. However, our study might be a starting point not to understand field-normalization as unquestioned standard, but to investigate this issue in future studies.

# 1    Introduction

In advanced bibliometrics, field-specific citation rates are targeted by applying field-normalized indicators in citation analyses (Waltman, 2016). The citation count of a focal paper is standardized based on the average citation count of papers in a reference set. Field-normalization is a standard concept in professional bibliometrics (Leydesdorff, Wouters, & Bornmann, 2016) which is not only widely used (Purkayasthaa, Palmaroa, Falk-Krzesinskib, & Baas, 2019), but also recommended as one of ten guiding principles for evaluation studies in the Leiden manifesto (Hicks, Wouters, Waltman, de Rijcke, & Rafols, 2015; Wilsdon et al., 2015). Much efforts in bibliometrics research is spent on various technical improvements in the calculation of field-normalized indicators (Wilsdon et al., 2015).

A list of possible reasons for field-specific differences in citation impact has been published by Waltman and van Eck (2013b): "Each field has its own publication, citation, and authorship practices, making it difficult to ensure the fairness of between-field comparisons. In some fields, researchers tend to publish a lot, often as part of larger collaborative teams. In other fields, collaboration takes place only at relatively small scales, usually involving no more than a few researchers, and the average publication output per researcher is significantly lower. Also, in some fields, publications tend to have long reference lists, with many references to recent work. In other fields, reference lists may be much shorter, or they may point mainly to older work. In the latter fields, publications on average will receive only a relatively small number of citations, while in the former fields, the average number of citations per publication will be much larger" (p. 833). Similar points are mentioned by Abramo, Cicero, and D'Angelo (2011). According to Wang, Song, and Barabási (2013), "comparison of different papers is confounded by incompatible publication, citation, and/or acknowledgment traditions of different disciplines and journals" (p. 127).

Field-specific impact differences have been published in many bibliometric studies. However, we could not find any study which asked whether there is a *causal link* between field assignment and citation impact of papers. A possible answer is that a field is nothing more than the differences in field-related attributes, i.e., there is no causal effect of the field on citations beyond these attributes. We would like to name this possible answer as "redundancy hypothesis". For example, a field could be characterized by a certain number of review papers, a certain number of co-authors, and the fact that most articles come from a certain geographical region (e.g., locally relevant research). If all these characteristics which confound the causal field effect were kept constant or equal across fields, the fields probably would no longer differ in average citation rates. The field effect is then nothing more than the effect of the field related attributes. The alternative answer is that a field is more than the effects of its attributes usually associated with fields. We call this possible answer the "the whole is more than its parts hypothesis". Even if the essential attributes of fields were kept constant across the fields, differences in citations would remain. In this case, field-normalization would be a prerequisite for citation analysis and cannot be replaced by the consideration of a set of field-related attributes in citation analyses.

The above mentioned reasons by Waltman and van Eck (2013b) for explaining field-specific citation rates point to possible confounding factors that are specific for both citation impact and field assignment. For example, one reason for field-specific differences of citation rates are – according to Waltman and van Eck (2013b) – differences in the number of co-authors. One simple reason for the correlation between the number of co-authors and the number of citations might be that research in some fields is more complex than in other fields and more co-authors lead to an increasing number of self-citations then. Since both field and number of co-authors might have an influence on citations, the relationship between field and citation impact could be confounded by the number of co-authors (see Pearl & Mackenzie, 2018). Using a small world example, we would like to demonstrate this possible confounding.

The small world example in Table 1 reveals that the number of authors might cause differences in citation rates which are also reflected in field-specific citation rates. Based on the data in the table, field 1 (m=60) has a lower citation rate than field 2 (m=90). Thus, one could have the (false) impression that field specifics cause different citation rates (i.e., the causal effect or mean difference of 30 citations between both fields). However, the view on the number of co-authors demonstrates that it is actually this variable which determines differences in mean citations: given the group of papers with 1 co-author the mean difference between the two fields is zero ($\Delta$=50-50=0). The same applies for the group of papers with 5 co-authors ($\Delta$=100-100=0). The overall causal effect of field irrespective of the number of co-authors is the weighted average of the mean differences of the two co-authorship groups. The weights are the marginal frequencies of the two co-authorship groups, i.e. 0.5. The causal field effect is zero as well. The difference between the prima-facie effect of 30 citations and the causal zero effect results from the correlation between the number of co-authors and field ($\varphi$=.60). Whereas field 1 is characterized by a low number of co-authors, field 2 has a high number. The columns "Propensity score" and "w" in Table 1 will be discussed (and explained) in the following.

Table 1. Small world example which demonstrates that citations might be more dependent on the number of co-authors than on the field

| Paper | Number of co-authors | Field | Citations | Propensity score p | w=1/p |
|---|---|---|---|---|---|
| Paper 1 | 1 | 1 | 40 | .80 | 1.25 |
| Paper 2 | 1 | 1 | 60 | .80 | 1.25 |
| Paper 3 | 1 | 1 | 40 | .80 | 1.25 |
| Paper 4 | 1 | 1 | 60 | .80 | 1.25 |
| Paper 5 | 1 | 2 | 50 | .20 | 5.00 |
| Paper 6 | 5 | 1 | 100 | .20 | 5.00 |
| Paper 7 | 5 | 2 | 150 | .80 | 1.25 |
| Paper 8 | 5 | 2 | 50 | .80 | 1.25 |
| Paper 9 | 5 | 2 | 150 | .80 | 1.25 |
| Paper 10 | 5 | 2 | 50 | .80 | 1.25 |

Many empirical studies in the field of bibliometrics have shown that not only the number of authors, but also many other factors are possibly related to citation impact differences between papers – besides their scientific quality. As factors potentially influencing citations (FICs) the number of pages, the reputation of the publishing journal, and many other factors have been identified (see overviews of FICs in Tahamtan & Bornmann, 2018, 2019; Tahamtan, Safipour Afshar, & Ahamdzadeh, 2016). Most of these FICs can also be used to characterize field-specific differences of papers. Having these many FICs in mind, it becomes suspicious that visible field effects in citation rates can be traced back to underlying factors which (at least partially) cause these differences. Since we empirically know that different numbers of co-authors are related to differences in citation rates, it might be the case that fields have different citation rates, because the average number of authors per field is different. Thus, the field effect is actually (partially) a co-author effect. The same might be true for any other FIC besides the number of co-authors (e.g., the number of pages).

In this study, we address the question how strong fields influence the citation rates of papers beyond FICs. We apply a variant of the family of procedures which can be subsumed under the term "propensity score matching" – an approach which has been introduced for measuring causal links – to test whether papers with the same characteristics (e.g., number of co-authors and number of pages) received different citation impact when they have been assigned to different fields. The fruitfulness of propensity score matching for bibliometric studies has been demonstrated by Mutz and Daniel (2012a), Mutz, Wolbring, and Daniel (2017) and Farys and Wolbring (2017). The results of the present study may have consequences for the approaches used in advanced bibliometrics: the necessity of field-normalization is justified, or it is questioned.

# 2 Field-differences in citation rates and normalization of citation impact in bibliometrics: a literature overview

The Leiden Manifesto for research metrics includes ten principles to guide research evaluation (Hicks et al., 2015). Principle six states that normalized citation impact indicators are required (instead of raw citation counts) because publication and citation practices vary by field (see also Wilsdon et al., 2015). Field-normalized indicators have been developed to enable comparisons between the citation impact of publications from different fields (Mingers & Leydesdorff, 2015; Waltman, 2016). These comparisons are necessary if units are evaluated publishing in different fields (e.g., research groups or universities). One of the most important field-normalized indicators is the mean-normalized citation score which relates the citation impact of a unit to a worldwide, field-specific reference value (Waltman, van Eck, van Leeuwen, Visser, & van Raan, 2011). Percentile-based indicators focus on the number or proportion of publications (of a unit) that belong to the top 10% or the top 1% of their field (Bornmann & Williams, in press). Waltman et al. (2012) regard the top 10% indicator "as the most important impact indicator in the Leiden Ranking" (p. 2425). Mutz and Daniel (2019) have recently suggested a statistical approach to model raw citations but considering reference values in the statistical model.

One of the most frequently discussed topics in bibliometrics related to field-normalization concerns the delineation of fields. The boundaries of fields can be defined and implemented in field-normalization in various ways. It seems impossible to verify satisfactorily that one solution is better suited than another for field-normalization. Most bibliometricians use multi-disciplinary classification systems as the basis for building reference sets which have a broad coverage of publications from various disciplines. Among these systems, the most popular are the subject categories based on sets of journals provided by Clarivate Analytics in the Web of Science (WoS, about 250 journal-based subject categories), by Elsevier in Scopus (Wang & Waltman, 2016; Wouters et al., 2015) or by Science-Metrix (Archambault, Beauchesne, & Caruso, 2011) spanning nearly all fields (Sugimoto & Weingart, 2015). The WoS subject categories were aggregated by the OECD to six rather broad categories[1]. The use of journal sets for defining fields seems obvious because fields become mature when they introduce their first professional journal (Sugimoto & Weingart, 2015). However, the results by Wang and Waltman (2016) suggest that "WoS and especially Scopus tend to be too lenient in assigning journals to categories. A significant share of the journals in both databases, but especially in Scopus, seem to have assignments to too many categories" (p. 359).

If journal sets are used for field categorization, each paper belongs to the subject category (one or more) of its publishing journal. One of the problems with this field classification approach is the presence of multi-disciplinary journals (e.g., *Nature* or *Science*); the papers in these journals cannot be assigned to meaningful subject categories based on journal classification (Kronman, Gunnarsson, & Karlsson, 2010). The journal classification system is also stretched to its limits in the case of emerging or interdisciplinary fields, because the papers in these fields are usually published in a wide range of different journals (Strotmann & Zhao, 2010). Another problem might be the fact that many journal categories are too broad (Haddow & Noyons, 2013). They seem to cover multiple fields (each with their own citation practices) in a single category (van Eck, Waltman, van Raan, Klautz, & Peul, 2013).

Instead of journal-based field classification systems, publication-based field classification systems can be used for normalizing citations which might offer a more fine-grained representation of fields than the journal-based system (Waltman & van Eck, 2019). As an alternative to journal sets, Waltman and van Eck (2012) introduced a method for algorithmically constructing classification systems (ACCS). The method also leads to a multi-disciplinary classification system which is based on direct citation relations between single papers. This method already plays an important role in scientometrics, because the measuring of field-normalized citation impact in the Leiden Ranking is based on this approach (see https://www.leidenranking.com/information/fields). The results of Klavans and Boyack (2017) indicate that algorithmically constructed classification systems might be more accurate than systems based on other methods: "Using the assumption that higher concentrations of references denote more accurate clusters, direct citation thus provides a more accurate representation of the taxonomy of scientific and technical knowledge than either bibliographic coupling or co-citation" (p. 984).

---

[1] http://help.prod-incites.com/inCites2Live/filterValuesGroup/researchAreaSchema/oecdCategoryScheme.html

However, Leydesdorff and Milojević (2015) criticize the approach of algorithmically constructed classifications systems: "because these 'fields' are algorithmic artifacts, they cannot easily be named (as against numbered), and therefore cannot be validated" (p. 201). In a case study, Haunschild, Schier, Marx, and Bornmann (2018) concluded: "one possible interpretation of our results is that the cluster algorithm used to construct ACCS is not able to distinguish properly between scientific fields" (p. 445). Haunschild, Marx, French, and Bornmann (2018) compared field- and time-normalized indicators using three different classification schemes: (i) WoS journal sets, (ii) the ACCS proposed by Waltman and van Eck (2012), and (iii) an intellectually-based classification system by experts from the Chemical Abstracts Service (CAS) for 2,690,143 papers in the field of chemistry. They found that the normalized citation scores agree better when WoS journal sets and CAS classifications are used than when either of them is compared with normalized citation scores based on the ACCS proposed by Waltman and van Eck (2012). Remarks on other multi-disciplinary classification systems (which are not very popular in bibliometrics) can be found in Wang and Waltman (2016).

For field-normalization in a specific discipline and its related disciplines, mono-disciplinary classification systems can be used instead of multi-disciplinary systems. Mono-disciplinary systems are usually based on intellectual classifications of papers (Bornmann, Marx, & Barth, 2013). This means that human indexers classify the publications by using a detailed and controlled vocabulary. Either experts in disciplines classify papers, as with papers in the Chemical Abstracts (CA) database (Bornmann & Daniel, 2008; Bornmann, Mutz, Neuhaus, & Daniel, 2008), or the authors of papers themselves undertake this exercise, as in economics with the Journal of Economic Literature (JEL) classification system (Bornmann & Wohlrabe, 2019). Several classification systems have already been used for field-normalization besides CA and JEL: Medical Subject Headings (MeSH) (Boyack, 2004), Physics and Astronomy Classification Scheme (PACS) (Radicchi & Castellano, 2011), and sections of the American Sociological Association (ASA) (Teplitskiy & Bakanic, 2016). Another advantage of mono-disciplinary classification systems is that the classification provides a relatively high level of detail (Wang & Waltman, 2016). It is, however, a disadvantage of these systems that they cannot be used for field-normalization exercises referring to a broad spectrum of disciplines.

The different approaches for assigning publications to fields which are used in bibliometrics to normalize citations are only vaguely connected to criteria for defining fields (Hammarfelt, 2018). Bibliometricians seem to prefer practically realizable approaches neglecting the criteria which Sugimoto and Weingart (2015) denote as (1) cognitive (shared bodies of content, theories, and methods), (2) social (communities of multiple researchers using specific terminologies or technical languages), and (3) institutional (institutionalization in the form of academic departmentalization). According to Sugimoto and Weingart (2015), "the maturation of a discipline is perhaps best represented by its institutionalization, particularly in the form of academic departmentalization" (p. 780). Andersen (2016) focusses on the cognitive and social dimensions in characterizing fields: "A scientific discipline (or specialty, field or domain) could be understood … as an epistemic unit consisting of a set of closely related cognitive resources such as, for example, concepts, models and theories, and as a social unit consisting of highly similar experts who were employing and at the same time developing their shared cognitive resources" (p. 2). A field can be defined as an "invisible college" (Ziman, 1996, p. 69) sharing a particular research tradition, having a "great man" (Sugimoto & Weingart, 2015, p. 782) in its history, and having access to specific governmental funding (e.g., specific programs for nanotechnology).

The basic reason for conducting field-normalization in bibliometrics is that impact should be measured without confounding factors: "ideally, one wants citation indicators to measure impact in a monotonic fashion: the higher the metric, the 'better' the paper" (Ioannidis, Boyack, & Wouters, 2016). In the definition of field-normalized indicators, many bibliometricians not only consider the field of papers, but also their publication years and document type. According to the document type, many studies published hitherto have found that "reviews" usually receive more citations than "articles" – both are the most important document types in bibliometric analyses (e.g., Lundberg, 2007). Besides field, time, and document type, many other FICs have been identified (e.g., number of authors, length of paper title, and number of cited references). Comprehensive overviews of these factors can be found in Tahamtan and Bornmann (2018), Hanel and Haase (2017), Onodera and Yoshikane (2014), and Didegah and Thelwall (2013).

The identification of FICs – as important influencing factors on citations besides cognitive influence – is rooted in the social-constructivist side of citation theories. This side stresses the rhetorical functions of citations and the impossibility to standardize citation behavior across scientists (Riviera, 2013): authors cite the same paper for different reasons and citations have different functions. For instance, Gilbert (1977) introduced the idea that referencing is an aid to persuasion of readers. The micro-sociological school which represents the social-constructivist theory tradition contends that "scientists may have complex citation motives that have not yet been clearly understood, and therefore, they cannot be satisfactorily described unidimensionally" (Liu, 1993, p. 13). Although all cited and citing works are related by the same link in citation indexes, the authors' motives and citation behavior can be different (Tahamtan & Bornmann, 2019).

The other side of citation theory is reserved by the normative theory of citation (Merton, 1973) which assumes that authors follow uniform standards in citing publications. According to this theory, citations can be used for evaluative purposes, since citers give credits to cited publications (or cited scientists) which have influenced their research. According to Merton (1988), "the reference serves both instrumental and symbolic functions in the transmission and enlargement of knowledge. Instrumentally, it tells us of work we may not have known before, some of which may hold further interest for us; symbolically, it registers in the enduring archives the intellectual property of the acknowledged source by providing a pellet of peer recognition of the knowledge claim, accepted or expressly rejected, that was made in that source" (p. 622).

Basically, both theories of the citation process try "to explain why author *x* cite[s] article *a* at time *t*" (Sandström, 2014, p. 59). Research suggests that both theories have their place in explaining citation behavior. The study by Judge, Cable, Colbert, and Rynes (2007) suggests that "universalistic, particularistic, and mixed universalistic-particularistic characteristics all play significant roles in the extent to which research articles in the field of management are cited" (p. 500). On one side, the results of the study question the normative theory of citation that the quality of papers is the main factor driving citations. On the other side, the study demonstrates the necessity to consider many FICs if one is intended to explain citation impact of papers.

# 3 Methods

## 3.1 Dataset used

We used the papers of the WoS from the Max Planck Society's in-house database – derived from the Science Citation Index Expanded (SCI-E), Social Sciences Citation Index (SSCI), and Arts and Humanities Citation Index (AHCI) provided by Clarivate Analytics (formerly the IP and Science business of Thomson Reuters). Citations were counted until the end of 2017. Since it was not feasible to base the statistical analyses on the complete in-house database, we drew a sample of 308,231 articles that were included in the analyses of this study. The procedure for drawing the sample followed several requirements.

We selected a sample covering a broad range of subject categories and as large as possible differences in citation counts between the subject categories. We exported the meta-data of all English papers and their citation counts of the document type "article" which were published between 2000 and 2005 and assigned to at least one of the following ten WoS subject categories: (1) Biodiversity conservation, (2) Computer science, Artificial intelligence, (3) Communication, (4) Engineering, petroleum, (5) Family studies, (6) Geriatrics & gerontology, (7) Immunology, (8) Physics, particles & fields, (9) Rehabilitation, and (10) Spectroscopy. Actually, we wanted to keep only those publications for which we had sufficient information about relevant meta-data, e.g., number of pages, number of co-authors, authors' countries, and Journal Impact Factor (JIF). This might provoke a selection effect in the sense that publications with complete metadata included in the analysis differ from publications with incomplete metadata not included in the analysis in the mean citations that affects the generalizability of the results.

There are many procedures to handle missing data in propensity score estimation (e.g., Malla et al., 2018). Given the large sample size of this study, we opted for the "Imputation With Constant Plus Missingness Indicators" (Cham & West, 2016, p. 431) instead of a multiple imputation procedure. In the first step, 0/1 indicator variables of the missing pattern of the covariates were created; in the second step, the missing values were replaced by an arbitrary value, here 0. In the third step, both the indicator variables and the completed covariates were included in the logistic regression to estimate the propensity scores. We assume missing at Random (MAR) (MAR, Mutz, Bornmann, & Daniel, 2015, p. 2327). Thus, the missingness is not random, but it can be explained by other covariates with complete information included in the analysis. Given these covariates, the missingness is random again, i.e., missingness cannot be explained anymore by any covariates.

The set of these WoS subject categories was identified as follows: The analyzed WoS subject categories should not have too low average citation rates, but they should differ as much as possible in their number of average citation rates. We selected each twentieth WoS subject category in the database ordered by their average citation rate in 2005 (starting with that category having the highest average citation rate) from those fulfilling the following criteria: (1) more than 100 papers of the document type "article" were published in 2005, and (2) only papers which belong to one of the following OECD categories are considered: Natural sciences, Engineering and technology, Medical and Health sciences, and Social sciences. Given the large sample size of this study and the relatively small proportion of publications with missing information, selection effects induced by the application of these criteria can be largely neglected.

### 3.2 Variables included

Table 2 shows the variables which we included in this study. The variable "total citation counts" is the metric which we would like to explain in the statistical analyses. As FICs, we included several variables and their two-way interactions in the statistical analyses. We tried to consider as many FICs as possible in our study. The overview of Tahamtan et al. (2016) revealed "28 factors, influencing the frequency of citations … which were classified into three categories: 'Paper related factors', 'Author related factors' and 'Journal related factors'" (p. 1198). However, we could not include in the present study all the many FICs proposed (investigated) in the past. The most important reason for the reduction was that many FICs must be preprocessed manually (human-based) and/or are in need of an analysis of the full-text of publications (e.g., the number of differential equations, the academic titles of the authors or the specific design of a study; see Antonakis, Bastardoz, Liu, & Schriesheim, 2014; Di Vaio, Waldenström, & Weisdorf, 2012; Robson & Mousquès, 2016). Other FICs cannot be considered, since reliable and valid information is not available for statistical analyses: for example, the phenomenon "obliteration by incorporation" (McCain, 2014) refers to "work that has become so accepted that it is no longer cited" (MacRoberts & MacRoberts, 2010, p. 2).

Table 2. Sample description (N=308,231 papers)

| No | Label | Missings (in percent) | Mean | Median | Standard deviation | Minimum | Maximum |
|---|---|---|---|---|---|---|---|
| 1 | Number of subject categories | 0 | 2.20 | 2 | 0.96 | 1 | 6 |
| 2 | Number of pages | 0.000013 | 9.80 | 8 | 7.80 | 0 | 878 |
| 3 | Number of co-authors | 0 | 5.45 | 4 | 18.97 | 1 | 942 |
| 4 | Number of author addresses | 0 | 3.32 | 3 | 3.19 | 2 | 115 |
| 5 | Number of joined countries | 0 | 1.30 | 1 | 0.87 | 1 | 22 |
| 6 | USA | 0 | 0.36 | 0 | 0.48 | 0 | 1 |
| 7 | Europe | 0 | 0.40 | 0 | 0.49 | 0 | 1 |
| 8 | Asia | 0 | 0.16 | 0 | 0.37 | 0 | 1 |
| 9 | Number of keywords | 0.21 | 5.21 | 5 | 4.09 | 0 | 10 |

| 10 | Number of title words | 0 | 11.99 | 11 | 5.09 | 1 | 61 |
| 11 | Number of cited references | 0 | 28.46 | 25 | 19.49 | 0 | 731 |
| 12 | Number of linked cited references | 0 | 19.45 | 15 | 16.89 | 0 | 596 |
| 13 | Journal Impact Factor | 0.08 | 2.26 | 1.47 | 2.55 | 0 | 46.23 |
| | Total citations | 0 | 30.08 | 13 | 96.12 | 0 | 22,415 |

Although it was not possible in this study to consider the complete set of FICs proposed in previous studies, we considered those factors for which (1) several previous studies exist which have investigated the FIC; (2) previous studies have already published meaningful relationships with citation counts; and (3) the data for calculating the FICs were available in the Max Planck Society's in-house database. For example, the tabulated overview of FICs published by Hanel and Haase (2017) shows that the JIF and the number of cited references – both FICs were included in this study – have been frequently investigated, see No 9 (JIF) and No 8 (number of cited references) in Table 2. The tabulated overviews by Onodera and Yoshikane (2014) and Didegah and Thelwall (2013) contain the additional important information for our FICs selection indicating whether the studies having investigated FICs report weak or strong relationships with citations or report contradictory results.

The first FIC in Table 2 is the **number of subject categories**. Papers often belong not only to one but several subject categories. The results by Chen, Arsenault, and Larivière (2015) show that "in each and every discipline – except Earth and Space –, the top 1% most cited papers exhibit higher levels of interdisciplinarity than papers in other citation rank classes" (p. 1043). Similar results are available from Larivière and Gingras (2010). One can expect that interdisciplinary papers attract attention and citations from more disciplines than mono-disciplinary papers. The "number of subject categories" was considered in this study in three ways. First, articles belonging to several of the selected 10 subject categories were randomly assigned to one of them (no duplicates). 2.48% of the papers included in the data analysis had duplicates. Secondly, the "number of subject categories" was included as a further covariate or FIC in the propensity score matching. Third, an analysis was performed for papers that were actually assigned to only one subject category (see the appendix).

The second FIC in Table 2 is the **number of pages** or the size of a publication. One might speculate that more extensive papers include more content (empirical results, ideas, conceptual explanations etc.) than papers with only a few pages which might lead to more citations to the content. In agreement with this assumption, the results of the early study by Gillmor (1975) show that in general the longer the paper, the more citations it received. The author investigated all paper published in *Journal of Atmospheric and Terrestrial Physics* between 1967 and 1973. Similar results have been published by Leimu and Koricheva (2005) based on ecological papers and by Stanek (2008) based on papers published in major astronomy journals. According to Falagas, Zarkali, Karageorgopoulos, Bardakas, and Mavros (2013) the correlation between paper length and citation numbers holds true (considering major general medicine journals) "after adjustment for several potentially confounding variables, such as the study design, prospective or retrospective nature of the study, abstract and title word count, number of author-affiliated institutions and number of bibliographic references".

The third FIC in Table 2 is the **number of co-authors** (having collaborated for a paper). Many studies have investigated the relationship between this FIC and the number of citations finding a positive correlation (e.g., Beaver, 2004; Benavent-Pérez, Gorraiz, Gumpenberger, & Moya-Anegón, 2012; Lawani, 1986; Tregenza, 2002). For ecology papers, the study by Leimu and Koricheva (2005) reveals that "papers with four or more authors received more citations than did papers with fewer authors" (p. 30). The proposed reasons for the dependence of the number of co-authors and citations are the multi-disciplinarity of papers with many differently skilled co-authors (Wray, 2006) or the benefits of labour division by bringing "complementary talents and expertise to a problem" (Walstad, 2002, p. 20). In addition, "the higher the number of authors, the larger the network of scientists that might know of one of them and, thus, cite them. Alternatively, the increase in citation rates with the number of authors might be related to an increased frequency of self-citations in the case of multi-authored papers" (Leimu & Koricheva, 2005, p. 30). However, the latter results could only be partly confirmed by Glänzel, Debackere, Thijs, and Schubert (2006). According to Valderas (2007), the longer the author list "the greater the probability of the paper being presented to several conferences is, especially if the team is multidisciplinary" (p. 1496) which might increase the number of citations.

The fourth FIC in Table 2 is related to the second: **number of author addresses**. One can expect that more co-authors for publications are related to more different author addresses (see Taborsky, 2009). For papers published in nanoscience and nanotechnology journals from 2007 to 2009, Didegah and Thelwall (2013) found a positive relationship between the number of institutions (mentioned on a paper) and the number of citations. The same has been reported for papers published by researchers affiliated with a Spanish university (Iribarren-Maestro, Lascurain-Sanchez, & Sanz-Casado, 2007). Although the number of co-authors and the number of author addresses might be positively correlated, both might not be related to citations in the same way: collaborations of authors between different institutions can lead to higher citation rates than collaborations within the same institutions (Whitfield, 2008). A similar result has been found by Dong, Ma, Tang, and Wang (2018) measuring innovation of research: "we discover striking phenomena where a smaller, yet more diverse team is more likely to generate highly innovative work than a relatively larger team within one institution" (see also Jones, Wuchty, & Uzzi, 2008).

The fifth FIC in Table 2 reflects a similar concept as the third and fourth FICs do: **number of joined countries**. The above mentioned study by Iribarren-Maestro et al. (2007) did not only find a positive correlation between the number of institutions and the citation counts, but also between the number of countries (indicated by the co-authors' addresses of a paper) and citation counts. Reasons for an increased citation counts of papers published in international collaborations might be the greater spreading of published papers and the greater probability of including talented people in a team of co-authors – less talented co-authors might have been found in the same country.

The sixth till eighth FICs in Table 2 also focus on countries, but target an effect other than the number of countries. We included the information in the statistical analysis, whether an author from the **USA, Europe** (EU28: Austria, Belgium, Bulgaria, Croatia, Cyprus, Czechia, Denmark, Estonia, Finland, France, Germany, Greece, Hungary, Ireland, Italy, Latvia, Lithuania, Luxembourg, Malta, the Netherlands, Poland, Portugal, Romania, Slovakia, Slovenia, Spain, Sweden, and United Kingdom), or **Asia** (China, Japan, South Korea, Taiwan, India, and Singapore), published the paper. These binary variables focus on possible national biases in citations (see Bornmann, Stefaner, de Moya Anegon, & Mutz, 2014; Bornmann, Wagner, & Leydesdorff, 2018; Lariviere, Gong, & Sugimoto, 2018). To avoid the overload of the statistical analyses with many redundant variables having no or scarce effects in the statistical analyses, a more coarse-grained country classification was used. Due to small numbers of articles in specific countries, there is the risk that the membership of certain fields can be clearly determined in the statistical analysis (probability of 1.0). This violates the assumption of propensity score matching that propensities are greater than 0 and less than 1.

According to the literature overview by Taborsky (2009) "regional affiliations" and "whether the authors are from native English speaking countries" (p. 105) are related factors with citations. The results by Pasterkamp, Rotmans, de Kleijn, and Borst (2007) show that "citation frequency was significantly augmented by nation oriented citation bias. This nation oriented citation behaviour seems to mainly influence the cumulative citation number for papers originating from the countries with a larger research output" (p. 153). Thus, one can expect papers with high citation impact from high research-performing countries. In other words, "international collaboration alone is not important, unless it is with a high impact nation" (Didegah & Thelwall, 2014, p. 174). The results of Gingras and Khelfaoui (2018) reveal that countries especially profit from collaborations with the USA.

The ninth FIC in Table 2 is the **number of key words** which are mentioned in a paper. This number might reflect not only the spread of topics a paper deals with, but also its inter-disciplinarity. More key words might indicate that topics from different fields are approached in a paper. The results by Chen et al. (2015) show that "specialties see their interdisciplinarity increase steadily with the increase of each percentile rank class" (p. 1042). Similar results have been published by Larivière and Gingras (2010) and Haustein, Larivière, and Börner (2014). In addition, more key words increase the probability of a paper to be discovered by a potential citer.

The tenth FIC in Table 2 is the number of words in the title of a paper (**number of title words**). This number of words might be referred to as one of many other "superficial factors that have nothing to do with the content [or quality] of the article" (Wesel, Wyatt, & Haaf, 2014, p. 1608). However, the results by Wesel et al. (2014) show that factors such as the number of title words can be correlated with citations (in certain fields). The results have been confirmed by other studies: the analysis by Letchford, Moat, and Preis (2015) "provides evidence that journals which publish papers with shorter titles receive more citations per paper. These results are consistent with the intriguing hypothesis that papers with shorter titles may be easier to understand, and hence attract more citations". A negative correlation was also found by Guo, Ma, Shi, and Zong (2018) in economics. However, the negative correlation was visible only for the publication years until 2000: "the results show that correlation between title length and the number of citations is negative between 1956 and 2000, but becomes positive after 2000, when online searches became the predominant method for literature retrieval" (p. 1531).

The eleventh and twelfth FIC in Table 2 refer to two variables: the **number of cited references** and the **number of linked cited references**. Both variables are highly correlated but can deviate from one another especially in fields with a low coverage of the literature in WoS (or any other database such as Scopus). Whereas the number of cited references counts all cited references in a paper, linked cited references refer to only that part of cited references which can be linked to a paper covered in the database (i.e., a source item). Several studies revealed that the citation count of papers is related to the number of references cited (see, e.g., Hegarty & Walton, 2012; Kostoff, 2007; Onodera & Yoshikane, 2014). Fok and Franses (2007) state that "longer articles with more references and also articles with more authors tend to get more citations" (p. 386). Webster, Jonason, and Schember (2009) found that "log citations and log references were positively related … In other words, reference counts explained 19% of the variance in citation counts" (p. 356). Webster et al. (2009) mention two reasons for the correlation of number of cited references and citation counts: "First, review articles (e.g., theoretical reviews, meta-analyses) tend to have more citations than and are cited more frequently than typical empirical articles. Second, scientists are humans, and humans crave recognition for their work and often participate in reciprocal altruism" (p. 349).

As the results of Ahlgren, Colliander, and Sjögårde (2018) show not only the number of cited references, but also the share of cited references to papers within WoS is related to citation counts which justifies the consideration of both in the present study: the number of references and the number of linked references. Thus, the coverage of databases influences the times papers get cited (Ioannidis et al., 2016): if the coverage is low in a specific field, few links can be established between cited references and citing papers leading to low citation counts. Thus, one might imagine that if the complete literature from all fields would be included in a database and all cited references could be linked to a source item, field-specific differences in citation rates might be reduced.

The thirteenth FIC in Table 2 is the **JIF**. One might oppose that the consideration of the JIF is a circular reasoning, since the JIF is itself based on citation counts. Previous studies have shown, however, that the JIF reflecting the reputation of journals is one of the most important FIC (see the FICs overview in Onodera & Yoshikane, 2014). The importance can be explained by the fact that "the more highly regarded a journal, the more likely it is that researchers will want to make use of its contents" (Meadows, 1998, p. 165). From the many studies investigating the JIF in the past, only a selected number can be presented here. Stanbrook and Redelmeier (2005) conclude that "journals act as passive conduits for scientific content, the impact of an article was strongly influenced by which journal published it. The journal impact factor appears to represent an accurate estimate of the relative ability of a journal to enhance article impact beyond the baseline contribution from article authors" (see also Knothe, 2006).

The results by Leimu and Koricheva (2005), Judge et al. (2007), Haslam and Koval (2010), and Ketzler and Zimmermann (2013) demonstrate the important role of the JIF as a FIC for management and ecological papers as well as papers in psychology and economics. Lariviere and Gingras (2010) conducted a quasi-experimental design to investigate the influence of the JIF on citations: they compared the citation impact of papers published twice in high and low impact journals (i.e., identical paper-pairs). The results show "that the journal in which papers are published have a strong influence on their citation rates, as duplicate papers published in high-impact journals obtain, on average, twice as many citations as their identical counterparts published in journals with lower impact factors" (p. 424). The results were confirmed by a study with a similar design conducted by Perneger (2010).

Lozano, Larivière, and Gingras (2012) investigated the correlation between JIF and citations over time and revealed a time-dependent effect: "Throughout most of the 20th century, papers' citation rates were increasingly linked to their respective journals' IFs. However, since 1990, the advent of the digital age, the relation between IFs and paper citations has been weakening. This began first in physics, a field that was quick to make the transition into the electronic domain. Furthermore, since 1990 the overall proportion of highly cited papers coming from highly cited journals has been decreasing and, of these highly cited papers, the proportion not coming from highly cited journals has been increasing" (p. 2140). One might expect an increasing effect of the JIF on citations depending on the JIF values. That this is in fact the case was demonstrated by Milojević, Radicchi, and Bar-Ilan (2016): "The benefit of publishing in a journal with a higher IF value, instead of another with a lower one grows slowly as a function of the ratio of their IF values. For example, receiving more citations is granted in 90% of the cases only if the IF ratio is greater than 6, while targeting a journal with an IF 2 higher than another brings in marginal citation benefit".

### 3.3 Statistical analyses

A causal inference is only possible in a classical randomized control group experiment, in which the units can be interchangeably assigned to the individual treatment groups. If the units are the papers and the treatments are the fields, the same paper cannot be assigned to all fields which is a prerequisite for a treatment. However, since interdisciplinary and transdisciplinary research strongly blurs the boundaries of fields, papers have cross-field characteristics. This can be especially assumed when the focus is not on individual journals but on fields defined by a set of different journal-based categories (e.g., WoS subject categories). Thus, interchangeability of units can largely be presupposed which questions the use of fields (or journal sets) for normalizing citation counts in bibliometrics. Articles with comparable values in all FICs might be interchangeable between the subject categories.

However, it is not possible to carry out a randomized group experiment (see AlShebli, Rahwan, & Woon, 2018): papers cannot be randomly assigned to fields for investigating the effect of the field on citation impact. But statistical concepts for causal inference, especially the so-called Rubin Causal Model (RCM, Rubin & Thomas, 1996), offer possibilities to draw causal conclusions for non-experimental designs as well. These manifold concepts are summarized under the term "propensity score matching" (Austin, Grootendorst, & Anderson, 2007; Rosenbaum & Rubin, 1983; Stuart, 2010). The method has already been used in bibliometrics (e.g., Colugnati, Firpo, de Castro, Sepulveda, & Salles, 2014; Farys & Wolbring, 2017; Mutz & Daniel, 2012b, 2012c; Mutz et al., 2017).

The small world example in Table 1 considers only one confounding variable (the number of co-authors). In the real world, not only one confounding variable can be assumed, but a set of (field related) variables that correlate both with fields (as treatment) and citations (as dependent variable). This assumption leads to the use of propensity scores. A paper's propensity score is its probability of belonging to a specific field, given a set of confounding variables. In the small world example in Table 1, field 1 is characterized by a low number of co-authors compared to field 2. This means that papers with a low number of co-authors have a high probability of belonging to field 1 or have a high propensity score (e.g., 0.8). With a large set of confounding variables, the effect of these confounding variables can be reduced on propensity scores. These scores might be calculated by a logistic regression of a field variable on a set of confounding variables. The confounding variables have the property that articles from different fields with the same propensity score are balanced with respect to all confounding variables. In this situation, field differences dwindle regarding these variables, propensity scores of different fields overlap, and fields no longer differ in the confounding variables.

There exist several variants of the propensity scores matching approach. One variant uses the propensity scores for weighting: papers with a certain FICs combination are weighted higher in the analyses than papers with another combination. Different ratios, e.g., different ratios of high- and low impact journals in the subject categories (e.g., journals from the social sciences or life sciences), can be compensated by different weighting of the corresponding papers irrespective of their citations. Since this method is often used in study designs with more than two groups, we used it in this study (this study is based on papers with more than two subject categories). In Table 1, the propensity scores for field 1 and field 2 and the weights are presented for the small-world example. The weighted mean value as the sum of the weighted citations, divided by the sum of the weights (=10) is 75 for both fields, that is, the mean difference between the two fields is zero.

From a statistical point of view, the "propensity score approach" is founded in Rubin's potential outcome concept (Mutz et al., 2017, p. 2142). The best way for causal inference is to collect citations of the same article (in this situation, all variables are equal) both under the condition of field A and under the condition of field B. The citations in the two fields would then be the "potential outcomes" of the article. The difference between the citations is the individual causal effect of field A and B on citations. In statistical terms, for an experiment with one treatment and one control for each unit i (here: a single paper), an individual treatment effect (here: a field) $\alpha_i$ can be defined as the difference between what was observed if the unit i (i.e., a single paper) is exposed to treatment of the respective field ($Y_i(1)$), and what was observed if the same unit is exposed to control or all other fields ($Y_i(0)$): $\alpha_i = Y_i(1) - Y_i(0)$, where $Y_i(1)$ und $Y_i(0)$ are the two potential outcomes for unit i. The fundamental problem of causal inference is that the individual causal effect $\alpha_i$ is in principle not possible to estimate, as in an observational study only one potential outcome of a unit is realized through the unit treatment assignment T (0 or 1) (Holland, 1986).

In a classical experiment, this problem is tackled by randomly assigning the units to the groups of interest. Individual effects for each single unit cannot be estimated in this setting, because only one potential outcome is realized in an experiment. However, at least the average effect between treatment and control can be thought of as an expected value, E, (i.e., the mean of the individual effects between treatment and control across all units). This value can be interpreted causally, as average causal effect (ACE):

$$ACE = E(Y_i(0)) - E(Y_i(1)) = E(Y_i(0|T=0)) - E(Y_i(1|T=1)), \qquad (1)$$

where $E(Y_i(0))$ and $E(Y_i(1))$ are the expected values under treatment and control. $E(Y_i(0|T=0))$ is the expected value of the potential outcome realized in treatment T=0 given the fact that paper i is in treatment (subject category) 0 and vice versa for $E(Y_i(1|T=1))$. In this case two potential outcomes are assumed for each unit, i.e. paper.

With non-experimental designs, however, selection effects are to be expected, i.e. the assignment to the groups is not at random. The general goal of this study is to model the selection or assignment process itself, i.e., to predict the probability of belonging to the treatment (here: fields) by means of a multiple logistic regression on an as large as possible number of covariates. The assignment probability is nothing else than the propensity score for a unit (here: a single paper) to belong to a certain treatment. Units from treatment and control with the same propensity score have comparable values in their covariates (**C**). This makes it possible to match treatment and control groups in order to arrive at groups with a balanced distribution of the covariates (resulting in an equivalent to a randomized control group experiment).

Three assumptions are of central importance in this respect (Mutz et al., 2017, p. 2143): (1) According to the "strong ignorability condition" the potential outcomes should be independent of the treatment (i.e., fields) given the covariates. This means that the specific combination of FICs of a paper can be ignored; the only decisive factor is the field to which the paper was assigned. (2) The propensity scores should be greater than zero and less than 1. For example, if a propensity score is 1.0, then a paper is uniquely assigned to a field based on its combination of covariates, but it cannot potentially belong to other fields. A comparison among treatments, i.e., subject categories, is no longer possible. This assumption can be checked by plotting the estimated propensity scores. In view of the huge sample size in this study, papers with propensity scores of 0 or 1 should also be considered. Thus, we set the propensity scores to 0.001 or 0.999. (3) Of less importance for this study is the so-called "stable unit value assumption" which excludes mutual influences of papers within a field. Although this assumption is violated in bibliometrics to some extent, we neglected it in view of the large data set used here.

The present study considers more than two treatment groups, so that propensity score approaches for more than two groups have to be applied (Imai & van Dyk, 2004; Lopez & Gutman, 2017; McCaffrey et al., 2013; Mutz & Daniel, 2012b). The basis of the estimation of propensity scores for multiple treatments is a multinomial multiple regression as an extension of a logistic regression to more than two groups. Instead of the usual group matching with regard to the propensity score, the "inverse-probability of treatment weighting" (IPW, McCaffrey et al., 2013, p. 3393) approach is used in this study. In this approach, the citation count of a unit (here: paper) is weighted with the inverse of the corresponding propensity score of the unit in a field. While the mean value per field results in an unconfounded mean value of that field, the differences of the weighted mean values result in an average causal effect between the two fields (McCaffrey et al., 2013, p. 3393):

$$\mu_t = \frac{\sum_{i=1}^{N} T_i[t] Y_i \omega_i[t]}{\sum_{i=1}^{N} T_i[t] \omega_i[t]} \qquad (2)$$

with weights $\omega_i$: $\omega_i[t] = \frac{1}{p_{ti}(\mathbf{C}_i)}$,

where $\mu_t$ is the unconfounded average of treatment or field t and $T_i[t]$ is the treatment or field assignment variable. The variable is 1, if the respective paper belongs to field t and 0 if not. The average causal effect (ACE) of field 1 to field 2 is the estimated mean difference $\mu_1-\mu_2$.

The suggested approach results in the following requirements for the selection of covariates (see section 3.2):

- The covariates should influence both the treatment assignment and the citation counts.
- The covariates should not lead to complete separation of the treatments or fields (i.e., the propensity score should greater than 0 and lower than 1.0).
- After the weighting with the propensity scores, the differences between the fields in the covariates should vanish. The fields should be similar regarding all covariates.
- A large set of covariates and their interactions should be used in order to guarantee comparability of the fields.

The selected approach and the corresponding requirements result in the following procedure (Spreeuwenberg et al., 2010, p. 167):

- Effect estimation before adjustment.
- Balance check before adjustment: it is checked, whether the fields – as treatments – differ in the set of covariates.
- Estimation of propensity scores: the propensity scores are estimated by a multinomial regression of citation counts on a set of covariates and their two-way interactions.
- Check for overlap: it is checked, whether the distributions of scientific impact between disciplines overlap (diagnostic I).
- Balance check after adjustment (diagnostic II).
- Estimation of the causal effects after adjustment.

The statistical analyses in this study were carried out with SAS 9.4 (SAS Institute Inc., 2012). The SAS program was checked on an example data set. The test revealed that the program was able to reveal the true unconfounded mean differences between groups, i.e. the average causal effects.

# 4  Results

The results section starts with reporting the diagnostic findings (in sections 4.1 and 4.2). Then, the estimated effects of the causal analyses – as the main results of the current study – are presented (in section 4.3).

## 4.1  Diagnostic I: assessing overlap of distributions

For a consistent estimation of the unconfounded field-specific effects, there should be an overlap between the fields regarding the propensity scores. In principle, each paper should be able to occur in all fields, a complete separation of the groups should not be possible (see McCaffrey et al., 2013, p. 3406). Since the propensity scores, i.e. the probabilities of paper allocations to all ten fields, are available for each paper of a particular field in this study, ten distributions can be represented for each field. Figure 1 shows the distribution of propensity scores for the ten WoS categories considered in this study as box plots. While the length of the boxes represents the interquartile distance (25% quartile to 75% quartile), the middle line in the boxes represents the median.

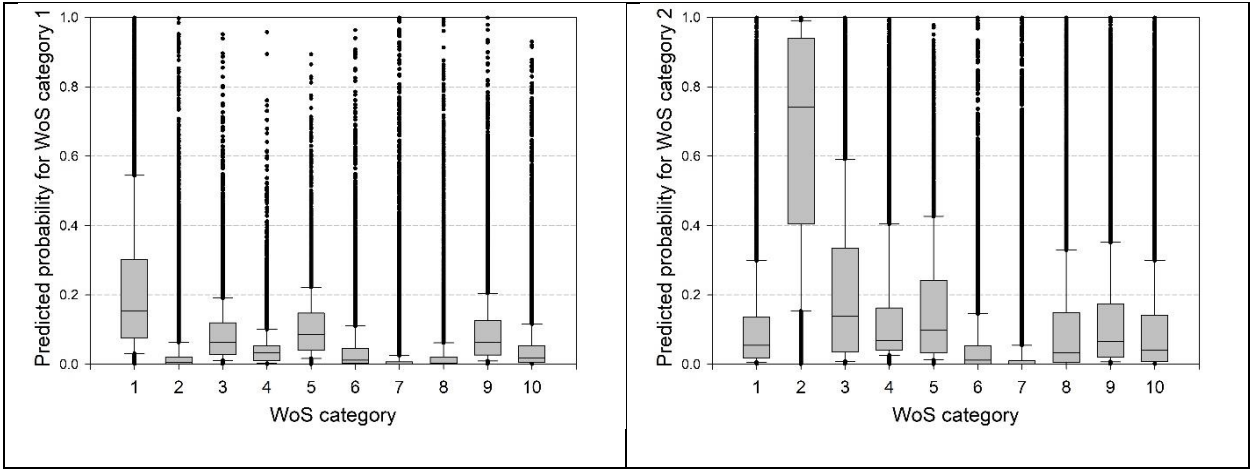

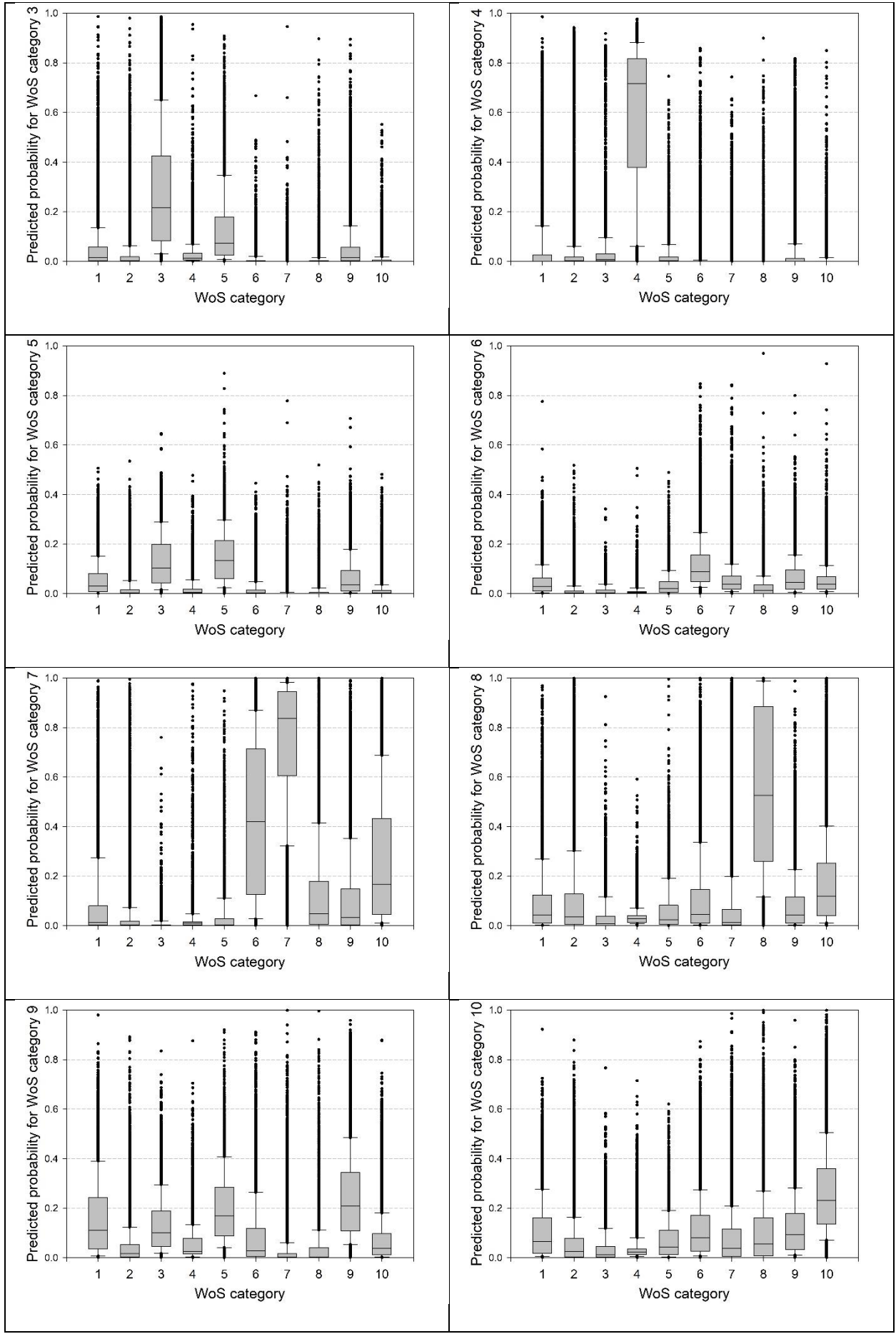

Figure 1. Boxplots for overlaps of multiple propensity scores (predicted probability) between the ten WoS categories included in this study (1=Biodiversity conservation, 2=Computer science, Artificial intelligence, 3=Communication, 4=Engineering, petroleum, 5=Family studies, 6=Geriatrics & gerontology, 7=Immunology, 8=Physics, particles & fields, 9=Rehabilitation, and 10=Spectroscopy)

It becomes clear in Figure 1 that the propensity scores – on average (median) – are highest for the fields to which the papers were assigned. For example, the second WoS category (Computer science, artificial intelligence) has the highest average propensity score for the WoS category 2 (highest picture on the right side of the figure). However, it also becomes clear for WoS category 2 that there is a greater than zero probability of potentially being able to be assigned to other WoS categories as well. Similar results as for WoS category 2 are observable for all WoS categories in

Thus, the above-mentioned assumption of overlapping distributions is more or less fulfilled for the dataset at hand.

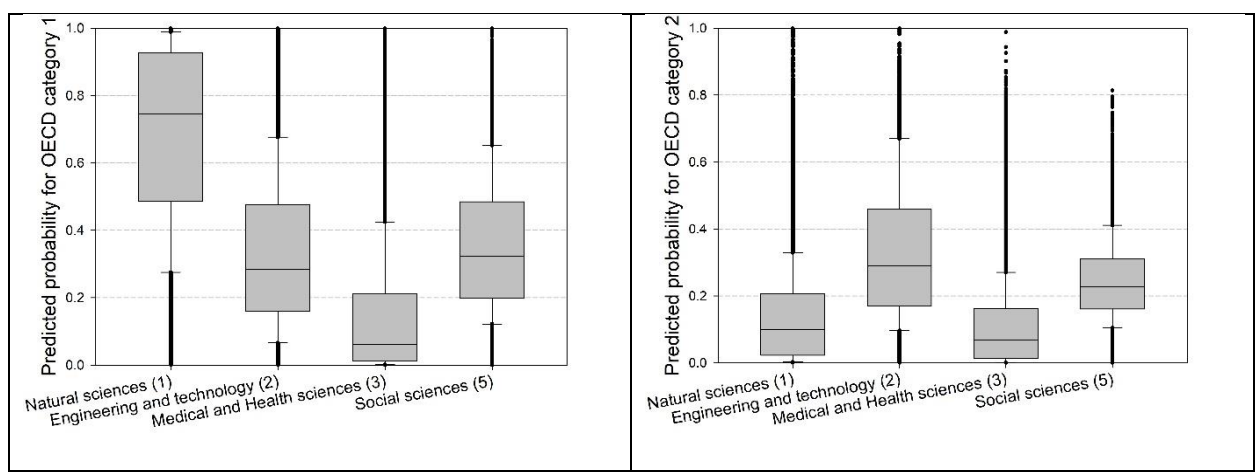

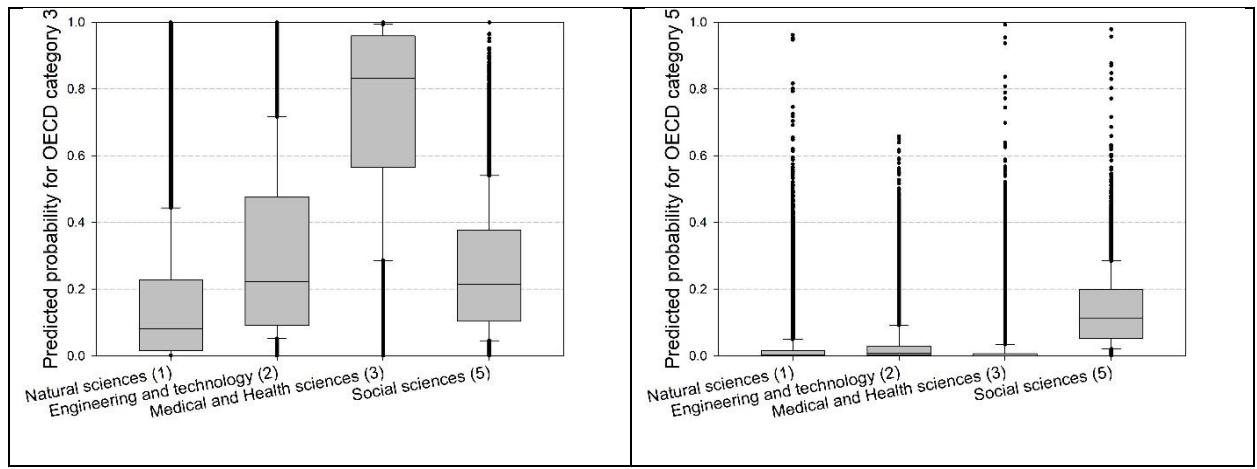

Figure 2. Boxplots for overlaps of multiple propensity scores between four OECD subject categories (papers published in the fourth category "Agricultural sciences" are not considered in this study)

In order to investigate the robustness of our results, we checked whether they changed when fields on a higher aggregation level than the WoS subject categories are considered in the statistical analyses. Although the OECD subject categories are based on the WoS subject categories, the composition of the fields varies by the sheer size. The OECD assigns all WoS subject categories to main disciplinary categories. The results based on these categories are presented in Figure 2. It becomes clear that the interpretation which we delivered based on the WoS subject categories holds true for OECD subject categories. Although the propensity scores are on a high average level for the category the paper belongs to, there is also considerable overlap of distributions.

## 4.2  Diagnostic II: balance check

The next diagnostic step was taken to examine whether the differences between the fields regarding the FICs vanish when the propensity scores are considered as covariates in a regression. For the analysis of the count variables (e.g., number of pages) a negative binomial regression was calculated; for the analysis of the binary variables (e.g., a paper published by a European author or not) the logistic regression was used, and for the approximately normally distributed variable, number of subject categories a papers belongs to, a normal distribution was assumed. The fields were included as additional covariates (as independent variables) in the model (ten WoS subject categories and four OECD subject categories, respectively) (Austin, 2009; Spreeuwenberg et al., 2010). In other words, each model included a single FIC (e.g., number of pages) as dependent variable and the propensity scores and subject categories as independent variables. Mean values for the FICs separated by subject categories were calculated both before and after adjusting for the propensity scores (see Table 3).

Table 3. Propensity score check of covariates for WoS subject categories

| Variable | Propensity score? | F-Test WoS (F(9, 308E3)) | Mean values – WoS subject categories | | | | | | | | | |
|---|---|---|---|---|---|---|---|---|---|---|---|---|
| | | | Biodiversity conservation | Communication | Computer science, artificial intelligence | Engineering, petroleum | Family studies | Geriatrics & gerontology | Immunology | Physics, particles & fields | Rehabilitation | Spectroscopy |
| Number of subject categories | No | 341.6 | 2.3 | 2.2 | 1.9 | 2.4 | 2.2 | 2.2 | 2.2 | 2.1 | 2.1 | 2.3 |
| | Yes | 7.0 | 2.2 | 2.2 | 2.2 | 2.2 | 2.2 | 2.2 | 2.2 | 2.2 | 2.2 | 2.2 |
| Number of pages | No | 7,694.2 | 11.5 | 12.4 | 18.0 | 6.1 | 15.0 | 7.9 | 7.3 | 11.8 | 9.6 | 8.4 |
| | Yes | 40.0 | 9.3 | 9.5 | 9.2 | 8.9 | 9.2 | 9.0 | 9.0 | 8.9 | 8.9 | 9.1 |
| Number of author addresses | No | 2,477.2 | 3.0 | 2.6 | 2.5 | 2.3 | 2.7 | 3.6 | 3.7 | 4.0 | 3.2 | 3.0 |
| | Yes | 6.0 | 3.1 | 3.3 | 3.3 | 3.2 | 3.2 | 3.2 | 3.2 | 3.1 | 3.2 | 3.2 |
| Number of co-authors | No | 6,438.6 | 3.0 | 4.3 | 2.0 | 2.2 | 2.6 | 4.5 | 6.1 | 9.3 | 3.4 | 4.3 |
| | Yes | 47.4 | 4.5 | 4.7 | 4.6 | 4.4 | 4.4 | 4.5 | 4.1 | 4.1 | 4.4 | 4.5 |
| Number of cited references | No | 6,455.7 | 38.1 | 19.9 | 38.6 | 8.7 | 37.1 | 34.0 | 32.5 | 30.2 | 30.7 | 23.6 |
| | Yes | 42.7 | 26.4 | 27.0 | 25.4 | 24.4 | 25.1 | 25.7 | 25.5 | 25.3 | 24.9 | 25.2 |
| Number of linked cited references | No | 15,633.2 | 18.7 | 7.7 | 11.8 | 2.0 | 16.4 | 25.8 | 28.7 | 20.9 | 16.9 | 16.3 |
| | Yes | 35.6 | 14.9 | 15.3 | 14.2 | 14.4 | 14.0 | 14.7 | 14.7 | 14.4 | 14.0 | 14.0 |
| USA | No | 1,999.2 | .47 | .21 | .63 | .22 | .74 | .47 | .41 | .31 | .53 | .26 |
| | Yes | 12.2 | .36 | .38 | .32 | .34 | .31 | .36 | .36 | .35 | .34 | .36 |
| Europe | No | 1,069.2 | .31 | .44 | .21 | .11 | .12 | .37 | .41 | .49 | .27 | .44 |
| | Yes | 6.2 | .35 | .34 | .38 | .38 | .38 | .34 | .34 | .33 | .34 | .33 |
| Asia | No | 998.9 | .05 | .26 | .05 | .06 | .02 | .08 | .16 | .20 | .06 | .18 |
| | Yes | 6.1 | .07 | .06 | .07 | .07 | .07 | .07 | .07 | .07 | .07 | .07 |
| Number of joined countries | No | 747.3 | 1.3 | 1.2 | 1.1 | 1.1 | 1.1 | 1.2 | 1.3 | 1.7 | 1.1 | 1.3 |
| | Yes | 0.8 | 1.3 | 1.3 | 1.3 | 1.3 | 1.3 | 1.3 | 1.3 | 1.3 | 1.3 | 1.3 |
| Number of title words | No | 9,435.1 | 13.5 | 9.2 | 11.4 | 9.5 | 11.9 | 12.8 | 14.5 | 9.2 | 12.4 | 13.3 |
| | Yes | 30.1 | 11.7 | 11.9 | 111.7 | 11.8 | 11.6 | 11.5 | 11.4 | 11.4 | 11.5 | 11.4 |
| Number of key words | No | 14,894.73 | 5.6 | 1.6 | 3.2 | 0.7 | 4.8 | 7.1 | 7.9 | 4.8 | 5.0 | 5.3 |
| | Yes | 38.7 | 3.8 | 4.0 | 3.7 | 3.9 | 3.6 | 3.7 | 3.7 | 3.7 | 3.6 | 3.6 |
| Journal Impact Factor | No | 16,510.4 | 1.3 | 0.7 | 0.6 | 0.2 | 0.9 | 2.2 | 3.7 | 3.1 | 0.9 | 1.7 |

| | | Yes | 16.5 | 1.5 | 1.5 | 1.4 | 1.5 | 1.4 | 1.4 | 1.4 | 1.4 | 1.4 | 1.4 |

Both for the WoS categories (see Table 3) and for the OECD categories (see Table 4), a drastic but not complete reduction of the mean differences was observed, if the ten (WoS) or five propensity score variables (OECD) were included in the regression analyses (grey background). For example, the mean values for the JIF are between 0.2 ("Engineering, petroleum") and 3.7 ("Immunology") in Table 3. If the propensity scores are considered in the regression analyses, the mean JIFs in the fields vary only between 1.4 or 1.5.

The results in both tables can be interpreted also from the direct comparison of the F-tests before and after the adjustment. For example, the F-test for the regression analysis including "number of pages" reduces from 7,694.2 to 40.0, i.e. by 99.5% (see Table 3), if the propensity score variables are included as covariates in the analysis. Note that we waived statistical significance testing for the interpretation of the results in view of the rather large sample sizes (see section 3.1). All effects would be statistically significant anyway (df = 308,000).

The results of the F-tests and the comparisons of the mean values for the FICs reveal that the fields are balanced with respect to the set of covariates. However, they are not completely balanced. Although differences between the fields remain, it is remarkable how much the differences shrank after adjustment by the propensity scores: only 12 FICs and their interactions have been considered (and many more could be considered).

Table 4. Propensity score check of covariates for four OECD subject categories

| Variable | Propensity score? | F-Test OECD (F(3,294E3)) | Mean values – OECD subject categories ||||
|---|---|---|---|---|---|---|
| | | | Natural sciences | Engineering and technology | Medical and Health sciences | Social sciences |
| Number of subject categories | No | 164.1 | 2.2 | 2.3 | 2.2 | 2.2 |
| | Yes | 0.74 | 2.2 | 2.2 | 2.2 | 2.2 |
| Number of pages | No | 14,507.0 | 12.1 | 9.2 | 7.7 | 15.0 |
| | Yes | 61.7 | 9.3 | 9.3 | 9.1 | 9.8 |
| Number of author addresses | No | 2,336.0 | 3.2 | 2.8 | 3.6 | 2.7 |
| | Yes | 1.7 | 3.2 | 3.2 | 3.2 | 3.2 |
| Number of co-authors | No | 5,437.7 | 6.3 | 3.5 | 5.5 | 2.6 |
| | Yes | 35.3 | 5.1 | 4.8 | 5.0 | 4.9 |
| Number of cited references | No | 4,316.0 1 | 26.1 | 22.4 | 32.4 | 37.1 |
| | Yes | 38.9 | 27.5 | 26.6 | 26.6 | 27.8 |
| Number of linked cited references | No | 13,137.5 | 14.4 | 12.5 | 26.7 | 16.4 |
| | Yes | 27.9 | 17.1 | 16.4 | 16.5 | 17.2 |
| USA | No | 3,458.9 | .28 | .30 | .43 | .74 |
| | Yes | 34.1 | .38 | .37 | .35 | .33 |
| Europe | No | 1,250.2 | .45 | .34 | .39 | .12 |
| | Yes | 8.6 | .34 | .35 | .36 | .38 |
| Asia | No | 1,350.2 | .22 | .13 | .14 | .02 |
| | Yes | 6.6 | .06 | .07 | .07 | .08 |
| Number of joined countries | No | 517.1 | 1.4 | 1.2 | 1.3 | 1.1 |
| | Yes | 1.0 | 1.3 | 1.3 | 1.3 | 1.3 |
| Number of title words | No | 20,019.4 | 9.6 | 12.2 | 14.1 | 11.9 |
| | Yes | 19.3 | 11.7 | 11.6 | 11.6 | 11.7 |
| Number of key words | No | 15,906.5 | 3.4 | 4.0 | 7.4 | 4.8 |
| | Yes | 20.7 | 4.6 | 4.5 | 4.5 | 4.7 |
| Journal Impact Factor | No | 16,013.6 | 1.8 | 1.2 | 3.2 | 0.9 |
| | Yes | 7.3 | 1.8 | 1.8 | 1.8 | 1.9 |

## 4.3 Estimated effects with and without adjustment for WoS subject categories

In the last step of the statistical analysis, the causal effects were estimated by weighting the citation data according to the propensity scores (see Table 5). In order to calculate the weighted average for a subject category, the citations of papers assigned to a subject category were multiplied by the weights (Eq. 2). The weights were derived from the propensity scores and divided by the sum of the weights. While the mean value differences between the WoS subject categories represent the confounded effects before the adjustment (before IPW), the mean value differences after the adjustment (after IPW) represent the causal or unconfounded effects of the WoS subject categories, i.e., the "pure" effects of the WoS subject categories on citations. The "pure" effect relates to the condition that the fields do not differ any more with respect to the FICs included in the analyses (see Table 3).

In the interpretation of the results in Table 3, not only the mean value differences between the subject categories after IPW can be considered, but also the changes of the mean values before and after the weighting for single subject categories. On one side, the mean differences between the WoS subject categories in Table 3 become smaller due to the adjustment. On the other side, the differences in citation rates are far greater than can be explained by the considered FICs. Both results speak for the "whole is more than its parts hypothesis" which claims that a field is more than the effects of its field attributes (e.g., typical number of co-authors). For example, the mean differences for "Spectroscopy", "Immunology", and "Biodiversity" strongly shrink from maximal 13 citations to maximal 2 citations after the adjustment. The absolute mean difference between "Biodiversity" and "Computers Science, artificial intelligence" does not change significantly after the adjustment (~10 citations), but the sign of the effect changes from positive to negative.

While the subject categories "Rehabilitation" and "Communication" show no major changes in the mean values, the subject categories "Computer Science, artificial intelligence" and "Spectroscopy" have significantly higher citation impact values after adjustment. For the subject category "Computer Science, artificial intelligence", the citation impact value rises from 23.20 to 40.70; for the subject category "Spectroscopy" from 20.01 to 29.65. The increase in the citation impact values for these fields points out that the citation impact is undervalued, if the uncorrected or confounded citation data are used. Since the results in Table 5 for "Immunology" demonstrate a decrease in citation impact values by considering the weighting (from 42.2 to 31.7), the impact is overvalued in this case. Changes in mean values before and after the adjustment should be interpreted with caution, since the model actually only defines mean differences among subject categories before and after the adjustment.

Many papers in the WoS database are assigned to more than one subject category. That means there are possible dependencies in the dataset that might distort the empirical results. However, even if the same analysis is undertaken for papers which were assigned to only one subject category (see the results in Table A1 in the appendix), differences among the fields remain. The result also speaks for the "whole is more than its parts hypothesis".

Table 5. Effects of WoS subject categories on citations before and after IPW

| WoS subject category | | Before IPW | | After IPW | |
|---|---|---|---|---|---|
| | Number of papers | Mean | Standard deviation | Mean | Standard deviation |
| Biodiversity conservation | 11,762 | 33.52 | 56.18 | 30.90 | 60.82 |
| Computer Science, artificial intelligence | 57,338 | 23.20 | 180.38 | 40.70 | 159.87 |
| Communication | 6,882 | 24.54 | 42.83 | 25.56 | 43.91 |
| Engineering, petroleum | 10,699 | 4.20 | 13.07 | 18.40 | 27.99 |
| Family studies | 6443 | 28.30 | 53.63 | 21.1 | 35.12 |
| Geriatrics & gerontology | 12,852 | 37.10 | 79.84 | 30.33 | 55.40 |
| Immunology | 100,748 | 42.24 | 73.12 | 31.70 | 61.74 |
| Physics, particles & fields | 51,245 | 25.37 | 58.29 | 23.33 | 62.61 |
| Rehabilitation | 19,278 | 25.81 | 44.32 | 23.93 | 47.29 |
| Spectroscopy | 30,984 | 20.01 | 39.86 | 29.65 | 123.11 |

Note. For IPW not only a weighted mean, but also a weighted standard deviation was calculated, where the divisor is the sum of weights.

## 4.4 Estimated effects with and without adjustment for OECD categories

There are also differences with regard to the OECD categories (see Table 6). While there are no differences for the OECD Category "Natural sciences" before and after IPW, the OECD Category "Engineering and technology" is clearly undervalued in the observed citation data. The mean value rises from 17.20 (before IPW) to 30.13 (after IPW). We focus the interpretation on the differences between the groups (causal effects) instead of the absolute values (because mean differences are defined in the causal model as causal effects but not absolute values): for example, the difference between "Medical and Health Sciences" and "Natural sciences" decreases from 14 citations (39.36 vs. 25.13) before IPW to only about 2 citations (34.15 vs. 32.36) after IPW. This means that the observed differences between "Natural sciences" and "Medical and Health sciences" are strongly determined by the field-related attributes. Although the difference between both fields decreases by IPW, it does not vanish completely. The average citation counts of "Social sciences" is about 10 citations lower than the average citation counts of the other three OECD categories. This pattern reflecting the "whole is more than its parts hypothesis" is also visible for other OECD categories.

We restricted this analysis also to papers that are assigned to only one subject category (see above). The results are in Table A2 in the appendix. The results reveal that strong differences among the subject categories remain. In other words, the results based on the OECD subject categories also confirm the "whole is more than its parts hypothesis".

Table 6. Effects of OECD categories on total citations before and after IPW

| OECD Category | | Before IPW | | After IPW | |
|---|---|---|---|---|---|
| | Number of papers | Mean | Standard deviation | Mean | Standard deviation |
| Natural sciences | 120,345 | 25.13 | 131.40 | 32.36 | 103.94 |
| Engineering and technology | 48,565 | 17.20 | 36.90 | 30.13 | 120.45 |
| Medical and Health sciences | 132,878 | 39.36 | 70.63 | 34.15 | 72.45 |
| Social sciences | 6,443 | 28.30 | 53.63 | 22.00 | 36.97 |

Note. For IPW not only a weighted mean, but also a weighted standard deviation was calculated, where the divisor is the sum of weights.

# 5 Discussion

Despite the observed differences in citation counts between fields, the question remains how strong fields (however defined) influence the citation counts of papers. The question arises because of two reasons: (1) studies which have investigated reasons to cite or citation functions revealed that many FICs exist – the field is only one FIC besides many others, and the field might be confounded with these other FICs. (2) It is not clear how a field can be properly defined (separated) in bibliometrics. Fields can be invisible colleges (see above), organizational units, and common research topics (among other things).

Ioannidis et al. (2016) explain the problem of usual field-normalization in bibliometrics as follows: "a major challenge is how to define scientific fields for normalization. Fields have been categorized in the past on the basis of journals or library categories. Within-field citations are usually denser than between-field citations. However, no field is isolated, and between-field communication is increasingly common nowadays … In some areas, the boundaries between fields seem to become less distinct. Different categorizations may segregate science to anywhere between a dozen … to several thousands of disciplines or specialties … Fields may be defined a priori (e.g., based on the journals where research gets published) or dynamically (e.g., based on citing or cited papers or on the references of citing or cited papers) … The rationale is that citing papers consider cited papers relevant to their work, so they belong to the same field. Obviously, this is not true for all citations – e.g., some methods (statistical, laboratory, or other) can be used by multiple unrelated fields, and there are also substantive interdisciplinary references".

Thus, if it is not possible to properly delimit the papers belonging to a single field in bibliometrics, how can we decide that the FIC "field" should be considered in measuring citation impact? A possible valid assignment of papers to fields might diminish field-specific differences in impact scores which are observable to various extents based on the field-categorization schemes used by bibliometricians today. In contrast, the number of co-authors – another FIC besides the field – is clearly defined and realizable in publication data but is usually not considered in normalization approaches of citation impact. When we started this study, we were confronted with the problem of selecting a field-categorization scheme for this study. Journal sets are one of the most popular field-categorization schemes for field-normalization. Thus, we took this scheme as basis for investigating the causal effects of fields on citations in the current study. In order to distill the "pure" effect of the field (defined by the scheme) on citation impact, this contribution is devoted to statistical concepts of causal inference.

Using the Max Planck Society's in-house version of the WoS database, we selected ten WoS subject categories (fields) which have as large as possible differences in mean citation rates. We were interested in the question whether the differences can be traced back to other FICs (than fields) or are "pure" field effects. We considered number of pages, number of co-authors, number of author addresses, number of joined countries, the binary national variables "Asia", "USA", and "Europe", number of keywords, number of title words, number of (linked) cited references, and JIF as FICs in this study. Although many more FICs have been proposed and investigated in previous studies, we included only a reduced set covering frequently studied FICs and FICs with a proposed significant influence on citation impact. However, the selection of the FICs might be a limitation of the study which is explained in the following (more FICs could have been included).

Although many studies have been conducted previously on FICs and the relationship between citation impact and fields, we are not aware of any other studies applying the propensity score matching approach being a family including many different procedures. This approach can be used for causal inferences comparable to an experimental design, however, with less methodological stringency due to the non-experimental nature of the used data. Most of the studies on FICs and the relationship between citation impact and fields were based on correlations or ordinary regression analyses which cannot make any claim to causality. Propensity score matching – a family of different procedures (propensity score matching, weighting, two groups, and multiple groups) – and the corresponding concept of causal inference, called Rubin Causal Model, defines statistical conditions being fulfilled to make causal statements. In propensity score matching (i.e. inverse propensity weighting) groups (here: fields) are established which do not differ anymore in central characteristics except for field. Very convincing for the advantage of propensity score matching are studies comparing the quasi-experimental design with randomized control group experiments on the same topic. The results of these studies revealed that the treatment effects did not differ when a quasi-experimental design with propensity score matching was compared with a real experiment (e.g., Cook, Shadish, & Wong, 2008).

In one of two diagnostic steps in our statistical analyses, we considered the calculated propensity scores as covariates in regression analyses to examine whether the differences between the fields vanish. As the results showed, the differences did not completely vanish, but were drastically reduced in most of the FICs up to 99% of the F-test value (before the adjustment). We received similar results when we calculated mean value differences of the fields after IPW representing the causal or unconfounded field effects on citations: the fields still differ with respect to the FICs included in the analyses.

Our results can be interpreted in different ways. One possible interpretation is that field-differences in citations exist which are relatively independent of other FICs. In this case, normalization of citation impact would be reasonable although our results might question the focus on fields in normalization approaches. Other FICs have (significant) effects on citation impact; a result which has been confirmed by other studies (e.g., Yu, Yu, Li, & Wang, 2014). Another possible interpretation of our results is that field-normalization does not only capture pure field effects but also other effects (FICs-related) which are desirable to include in normalization procedures. Field-differences could vanish in principle, e.g., by including many more FICs in the analysis than we did in this study.

Although the results of our study gain important insights in the relation of field-normalization and FICs (field-related attributes of papers), future studies could build upon this study by using an improved design to allow more pronounced interpretations. The following five potential limitations of this study prevented a clear-cut result and therefore should be used to improve the design of future studies:

- *Selection of covariates:* For the propensity score matching approach, an as large as possible number of covariates is required to establish comparability of the groups (here: fields). We included important covariates in this study, but did not consider many other covariates proposed in the past. For example, covariates explicitly referring to the quality of papers – independent of citations – were not available for our data at hand (e.g., quality assessments by peers). Since the normative theory of citing introduced by Merton (1973) bases citation decisions exclusively on the quality concept, an important factor is not considered in this study.

  - *Missing metadata*: There are missing values in some covariates. A missing imputation procedure was applied to avoid selection effects, but data analysis with missing imputed data is not fully efficient as data analysis with complete metadata.

- *Balance of the groups*: Although a strong reduction of the mean differences between the fields in the covariates could be observed, the differences do not vanish completely. Given the rather large sample size in this study, this is not surprising: more papers usually mean a greater heterogeneity among the papers which might be more difficult to capture with the set of covariates considered in this study. The consideration of more covariates might be a possible solution for this problem (in future studies). It cannot be excluded in this analysis and, particularly, in analyses with greater sets of covariates, however, that the assignment or not-assignment of papers to subject categories can be perfectly predicted (propensity score of 1 or 0). This violates the assumption that propensity scores should greater than 0 and less than 1.

- *Field-categorization scheme*: We used WoS subject categories as field-categorization scheme in this study. In section 2, we outlined that several schemes exist in bibliometrics (and no standard scheme). Since the schemes are conceptually very different, a considerable impact of the schemes on field-specific impact scores might be possible. Thus, it would be interesting to include not only one scheme – as we did – but several schemes in future studies (e.g., based on citation relations or human-based field-assignments). Then, the influence of the schemes – independent of the field – could be estimated on the results.

- *Set of subject categories*: We used a specific set of subject categories arbitrarily chosen to cover a broad range of fields. Strictly speaking, the results of this study are restricted to these subject categories.

Although the results of our study point out that field-effects on citations still exist even if FICs are considered, the reduction of field-effects might question the use of field-categorization schemes for normalizing citations. Since citation impact can be alternatively measured in research evaluation, these alternatives might be preferred. Two alternatives have been proposed in previous years which allow a kind of benchmarking "that is needed to enhance comparability across diverse scientists, fields, papers, time periods, and so forth" (Ioannidis et al., 2016):

(1) The first alternative are citing-side indicators (Zitt & Small, 2008) which focus on the basic premise of normalization that "not all citations are equal" (Ioannidis et al., 2016). In this study, we targeted the cited-side indicator's approach which started from the premise that not all papers are equal (but are embedded in a certain field-specific publication and citation culture). With citing-side indicators, every citation of a paper is normalized considering the citation density in the citing paper (measured on the basis of its cited references or its publishing journal). Thus, citing-side indicators are "normalised citation impact indicators without defining fields in an explicit way" (Wilsdon et al., 2015, p. 32). In recent years, several variants of citing-side indicators have been proposed. The results by Waltman and van Eck (2013a) show that they "may yield more accurate results" (p. 699) than usual field-normalized indicators. Bornmann and Marx (2015) partly confirmed the results. However, Bornmann and Haunschild (2017) found high correlations between cited-side and citing-side indicators.

(2) Another alternative to field-normalization in evaluative bibliometrics is to use non-normalized indicators and to "contextualize these indicators with additional information that enables evaluators to take into account the effect of field differences … For instance, to compare the productivity of researchers working in different fields, one could present non-normalized productivity indicators (e.g., total publication or citation counts during a certain time period) for each of the researchers to be compared. One could then contextualize these indicators by selecting for each researcher a number of relevant peers working in the same field and by also presenting the productivity indicators for these peers. In this way, each researcher's productivity can be assessed in the context of the productivity of a number of colleagues who have a reasonably similar scientific profile" (Waltman & van Eck, 2019, p. 295). The use of non-normalized indicators would also agree to the demand to use simple indicators in evaluative bibliometrics: "there is often a demand for simple measures because they are easier to use and can facilitate comparisons" (University of Waterloo Working Group on Bibliometrics, 2016, p. 2). These indicators are better understandable by experts in the evaluated fields and facilitate proper interpretation of the bibliometric results. A drawback of the contextualization approach is that it can scarcely be applied to researchers or institutions working in more than one field.

We would like to emphasize that we did not produce conclusive results for finally deciding on the use of field-normalized indicators in evaluative bibliometrics. Our study might be a starting point not to understand field-normalization as unquestioned standard, but to investigate this issue in future studies.

# Acknowledgements

The bibliometric data used in this study are from the bibliometric in-house databases of the Max Planck Society (MPG) and the Competence Centre for Bibliometrics (CCB, see: http://www.bibliometrie.info/). The MPG's database is developed and maintained in cooperation with the Max Planck Digital Library (MPDL, Munich); the CCB's database is developed and maintained by the cooperation of various German research organizations. Both databases are derived from the Science Citation Index Expanded (SCI-E), Social Sciences Citation Index (SSCI), Arts and Humanities Citation Index (AHCI) prepared by Clarivate Analytics, formerly the IP & Science business of Thomson Reuters (Philadelphia, Pennsylvania, USA).

# Appendix

Table A1. Effects of WoS subject categories on citations before and after IPW for papers **not** classified in other subject categories

| WoS subject category | | Before IPW | | After IPW | |
|---|---|---|---|---|---|
| | Number of papers | Mean | Standard deviation | Mean | Standard deviation |
| Biodiversity conservation | 511 | 13.30 | 18.16 | 15.76 | 18.48 |
| Computer Science, artificial intelligence | 17,274 | 19.94 | 242.34 | 28.58 | 214.03 |
| Communication | 2,340 | 24.69 | 44.48 | 17.72 | 30.82 |
| Engineering, petroleum | 1,124 | 7.71 | 19.17 | 16.10 | 29.35 |
| Family studies | 518 | 19.97 | 26.57 | 13.80 | 19.79 |
| Geriatrics & gerontology | 3,285 | 25.33 | 37.77 | 21.69 | 35.89 |
| Immunology | 24,184 | 61.38 | 100.84 | 32.68 | 65.48 |
| Physics, particles & fields | 15,274 | 26.52 | 69.87 | 37.61 | 289.81 |
| Rehabilitation | 3,737 | 20.88 | 29.25 | 20.33 | 30.98 |
| Spectroscopy | 7,781 | 18.74 | 38.50 | 22.79 | 93.62 |

Note. For IPW, not only a weighted mean, but also a weighted standard deviation was calculated, where the divisor is the sum of weights.

Table A2. Effects of OECD categories on total citations before and after IPW for papers **not** classified in other subject categories

| OECD category | Number of papers | Before IPW | | After IPW | |
|---|---|---|---|---|---|
| | | Mean | Standard deviation | Mean | Standard deviation |
| Natural sciences | 33,059 | 22.88 | 181.55 | 17.71 | 117.05 |
| Engineering and technology | 11,245 | 18.87 | 38.65 | 22.94 | 57.54 |
| Medical and Health sciences | 31,206 | 52.73 | 91.60 | 39.23 | 72.39 |
| Social sciences | 518 | 19.97 | 26.57 | 14.16 | 19.97 |

Note. For IPW, not only a weighted mean, but also a weighted standard deviation was calculated, where the divisor is the sum of weights.